\documentstyle[12pt,aasms]{article}
\begin{document}
\def\npix{{6068}}
\def\tcbr{{2.728 K}}
\def\tcbrerr{{2 mK}}
\def\trand{{10 $\mu$K}}
\def\rmspct{{50 ppm}}
\def\tcbrpm{{2.728 $\pm$ 0.004 K (95\% CL)}}
\def\adip{{3.369 mK}}
\def\dipdir{{$(\ell,b)=(264.14^\circ\pm0.15,48.26^\circ\pm0.15)$}}
\def\adippm{{$3.372 \pm 0.007$ mK (95\% CL)}}
\def\tdip{{2.717 K}}
\def\tdippm{{$2.717\pm 0.007$ K}}
\def\dipdev{{14 kJy/sr}}
\def\muval{{$-1\pm 4\times10^{-5}$}}
\def\mumin{{$9\times10^{-5}$}}
\def\yval{{$-1\pm 6\times10^{-6}$}}
\def\ymin{{$15\times10^{-6}$}}
\def\eg{{\it e.g. }}
\def\ie{{\it i.e. }}
\def\FIRAS{{FIRAS}}
\def\COBE{{\sl COBE\/}}
\def\icm{{cm$^{-1}$}}
\def\specpaper{Mather \etal\ (1994)}
\def\calpaper{Fixsen \etal\ (1994b)}
\def\dipolepaper{Fixsen \etal\ (1994a)}
\def\analysispaper{Wright \etal\ (1994)}

\hfuzz=10pt \overfullrule=0pt

\def\blankline{\par\vskip \baselineskip}

\def\Received{\par\vskip 36 pt
        \centerline{\it Received  \vrule height -1 pt width 2.5 in
        depth 1.8 pt .}\par\blankline}

\def\refitem #1! #2! #3! #4;{\hang\noindent
    \hangindent 20pt\rm #1, \it #2, \bf #3, \rm #4.\par}
\def\bookref{\par\noindent\hangindent 20pt}
 
\pretolerance=10000
\raggedright

\def\folio{\ifnum\pageno=1\nopagenumbers\else\number\pageno\fi}

\def\wisk#1{\ifmmode{#1}\else{$#1$}\fi}

\def\lt     {\wisk{<}}
\def\gt     {\wisk{>}}
\def\le     {\wisk{_<\atop^=}}
\def\ge     {\wisk{_>\atop^=}}
\def\lsim   {\wisk{_<\atop^{\sim}}}
\def\gsim   {\wisk{_>\atop^{\sim}}}
\def\kms    {\wisk{{\rm ~km~s^{-1}}}}
\def\Lsun   {\wisk{{\rm L_\odot}}}
\def\Msun   {\wisk{{\rm M_\odot}}}
\def\um     {\wisk{{\rm \mu m\ }}}
\def\sig    {\wisk{\sigma}}
\def\etal   {{\sl et~al.\ }}
\def\bsl    {\wisk{\backslash}}
\def\by     {\wisk{\times}}

\def\amin   {\wisk{^\prime\ }}
\def\asec   {\wisk{^{\prime\prime}\ }}
\def\cc     {\wisk{{\rm cm^{-3}\ }}}
\def\deg    {\wisk{^\circ\ }}
\def\ddeg   {\wisk{{\rlap.}^\circ}}
\def\damin  {\wisk{{\rlap.}^\prime}}
\def\dasec  {\wisk{{\rlap.}^{\prime\prime}}}
\def\approxeq{$\sim \over =$}
\def\abouteq{$\sim \over -$}
\def\percm{cm$^{-1}$}
\def\percmsq{cm$^{-2}$}
\def\percmcub{cm$^{-3}$}
\def\perhz{Hz$^{-1}$}
\def\perpc{$\rm pc^{-1}$}
\def\persec{s$^{-1}$}
\def\peryr{yr$^{-1}$}
\def\te{$\rm T_e$}
\def\tenup#1{10$^{#1}$}
\def\to{\wisk{\rightarrow}}
\def\thin{\thinspace}
%
%
%
\def\singlespace {\smallskipamount=3pt plus1pt minus1pt
                  \medskipamount=6pt plus2pt minus2pt
                  \bigskipamount=12pt plus4pt minus4pt
                  \normalbaselineskip=12pt plus0pt minus0pt
                  \normallineskip=1pt
                  \normallineskiplimit=0pt
                  \jot=3pt
                  {\def\smallskip {\vskip\smallskipamount}}
                  {\def\medskip   {\vskip\medskipamount}}
                  {\def\bigskip   {\vskip\bigskipamount}}
                  {\setbox\strutbox=\hbox{\vrule
                    height8.5pt depth3.5pt width 0pt}}
                  \parskip 0pt
                  \normalbaselines}
\def\doublespace {\smallskipamount=6pt plus2pt minus2pt
                  \medskipamount=12pt plus4pt minus4pt
                  \bigskipamount=24pt plus8pt minus8pt
                  \normalbaselineskip=24pt plus0pt minus0pt
                  \normallineskip=2pt
                  \normallineskiplimit=0pt
                  \jot=6pt
                  {\def\smallskip {\vskip\smallskipamount}}
                  {\def\medskip   {\vskip\medskipamount}}
                  {\def\bigskip   {\vskip\bigskipamount}}
                  {\setbox\strutbox=\hbox{\vrule
                    height17.0pt depth7.0pt width 0pt}}
                  \parskip 12.0pt
                  \normalbaselines}
%
%
\def\halfspace {\smallskipamount=6pt plus2pt minus2pt
                  \medskipamount=12pt plus4pt minus4pt
                  \bigskipamount=24pt plus8pt minus8pt
                  \normalbaselineskip=16pt plus0pt minus0pt
                  \normallineskip=2pt
                  \normallineskiplimit=0pt
                  \jot=6pt
                  {\def\smallskip {\vskip\smallskipamount}}
                  {\def\medskip   {\vskip\medskipamount}}
                  {\def\bigskip   {\vskip\bigskipamount}}
                  {\setbox\strutbox=\hbox{\vrule
                    height17.0pt depth7.0pt width 0pt}}
                  \parskip 12.0pt
                  \normalbaselines}

\def\pprintspace {\smallskipamount=4pt plus1pt minus1pt
                  \medskipamount=9pt plus2pt minus2pt
                  \bigskipamount=16pt plus4pt minus4pt
                  \normalbaselineskip=14pt plus0pt minus0pt
                  \normallineskip=1pt
                  \normallineskiplimit=0pt
                  \jot=4pt
                  {\def\smallskip {\vskip\smallskipamount}}
                  {\def\medskip   {\vskip\medskipamount}}
                  {\def\bigskip   {\vskip\bigskipamount}}
                  {\setbox\strutbox=\hbox{\vrule
                   height9.5pt depth4.5pt width 0pt}}
                  \parskip 0pt
                  \normalbaselines}
\def\reidelspace {\smallskipamount=.1667 true in plus4pt minus2pt
                  \medskipamount=.3333 true in plus8pt minus2pt
                  \bigskipamount=13 true pt plus2pt minus2pt
                  \normalbaselineskip=13 true pt plus0pt minus0pt
                  \normallineskip=1 true pt
                  \normallineskiplimit=0 true pt
                  \jot=3pt
                  {\def\smallskip {\vskip\smallskipamount}}
                  {\def\medskip   {\vskip\medskipamount}}
                  {\def\bigskip   {\vskip\bigskipamount}}
                  {\setbox\strutbox=\hbox{\vrule
                    height8.5pt depth3.5pt width 0pt}}
                  \parskip 0pt
                  \normalbaselines}

\pretolerance=1000	

\title{The Cosmic Microwave Background
Spectrum\\
from the Full {\sl COBE\footnotemark[1] FIRAS} Data Set
\footnotetext[1]{{\rm \doublespace 
The National Aeronautics and Space Administration/Goddard
Space Flight Center (NASA/GSFC) is responsible for the design, development,
and operation of the Cosmic Background Explorer ({\sl COBE}).
Scientific guidance is provided by the {\sl COBE} Science Working Group.
GSFC is also responsible for the development of the analysis software
and for the production of the mission data sets.}}
}
Submitted to the Astrophysical Journal \hfill {\tt astro-ph/9605054}

\author{
D.~J.~Fixsen\altaffilmark{1},
E.~S.~Cheng\altaffilmark{2},
J.~M.~Gales\altaffilmark{1}, 
J.~C.~Mather\altaffilmark{2}, 
R.~A.~Shafer\altaffilmark{2}, and 
E.~L.~Wright\altaffilmark{3} 
}

\altaffiltext{1}{Applied Research Corporation, Code 685.3, NASA/GSFC, 
Greenbelt MD 20771}
\altaffiltext{2}{NASA Goddard Space Flight Center, Code 685, Greenbelt MD 
20771}
\altaffiltext{3}{UCLA Physics and Astronomy, PO Box 951562, 
Los Angeles CA 90095-1562}

\noindent
\begin{abstract}

We have refined the analysis of the data from the \FIRAS\ (Far InfraRed Absolute
Spectrophotometer) on board the \COBE\ (COsmic Background Explorer).
The \FIRAS\ measures the difference between the cosmic microwave background
and a precise blackbody spectrum. We find new tighter upper limits on general 
deviations from a blackbody spectrum. The RMS deviations are less than 50 parts
per million of the peak of the CMBR. For the Comptonization
and chemical potential we find $|y|<$\ymin\ and $|\mu|<$\mumin\ (95\% CL). 
There are also refinements in the absolute temperature, \tcbrpm, 
and dipole direction, $(\ell,b)=(264.14^\circ\pm0.30,48.26^\circ\pm0.30)$
(95\% CL), and amplitude, \adippm.  All of these results 
agree with our previous publications.
\end{abstract}

\keywords{cosmology: cosmic microwave background --- cosmology: observations} 

\section{Introduction} 

The{~\FIRAS}~(Far~Infrared~Absolute~Spectrophotometer) instrument aboard the
\COBE\ (COsmic Background Explorer) satellite (Boggess \etal\ 1992 and
references therein; Bennett \etal\ 1992a) was designed to  measure the spectrum
of the cosmic microwave background radiation (CMBR). In the simple hot Big Bang 
model the spectrum has a blackbody form, but it could be distorted by energy 
release after a redshift $z \sim3\times10^6$ (Peebles 1971, Sunyaev \&  
Zel'dovich 1980). After the annihilation of positrons and the 
decoupling of neutrinos until $ z \sim 3 \times 10^4$, the CMBR was the 
dominant energy field.  The number of photons exceeds the number of baryons by 
a factor $\sim 10^9$, so excellent sensitivity is required to detect 
even large radiant energy releases.

Spectral distortion limits from the FIRAS were presented by  \specpaper.
These were based on $\sim 40$\ days of high Galactic latitude data from
a single detector and scan mode. The dipole spectrum was also determined 
(Fixsen \etal\ 1994a) from
this single detector.  \analysispaper\ found that the spectrum and dipole
are consistent with a simple Big Bang model.

All of the previous work in this frequency range
(2 to 21 \icm) was based on part of the data from a single detector.
This paper uses all of the low frequency data from the 10 month mission, 
except for the first month when settings were frequently readjusted to reach
the optimum condition, approximately doubling the statistical weight of the results.
There have been several improvements in the calibration 
that we note, but the basic calibration process remains the same.

The instrument was recalibrated using the method described in \calpaper. The
calibration was applied to the sky data, producing spectra from 2 to 21~\icm\ 
(5000 to 480 $\mu$m). Five important improvements were made to the calibration: 
(1) a bias in some pixels is corrected; (2) insufficiently sampled 
pixels ``borrow" data from neighboring pixels to determine a template for 
deglitching (removing effects of cosmic ray hits on the detector); 
(3) data with a large number of glitches (cosmic ray hits) are deweighted 
relative to data with few glitches;
(4) we use 320 points in the spectrum rather than 256; and (5) the data
are ``destriped" after the calibration.  Each of these processes is described
in more detail below.

\section{Instrument Design, Operation and Data}

The FIRAS is a polarizing Michelson interferometer, described by Mather 
\etal\ (1990, 1993). It measures the spectral difference between a 7\deg\ patch 
of sky and an internal blackbody.
The symmetric FIRAS optics are differential, with two input and two output
ports.  One input port receives emission from the sky,
defined by a non-imaging concentrator.
The other input port receives emission from an internal 
reference calibrator (emissivity $\approx 0.98$) with an associated concentrator. 
Each of the two output beams  (arbitrarily labeled ``left" and
``right") is split by a dichroic filter into low and high frequency beams, 
separated at 20 \icm, feeding four silicon composite bolometer 
detectors operated simultaneously.

An external blackbody calibrator provides the critical absolute calibration.  
The other elements are operated in the same manner while taking sky data or 
calibration data so the $only$ requirement on them is that they be repeatable.  
Although we verify that the emissivities of the concentrators and internal
calibrator are reasonable, the fundamental comparison is 
between the sky and the external calibrator. 
During calibration the sky aperture is completely filled by the external
calibrator with an emissivity $>0.99997$, calculated and measured.
The external calibrator is isothermal to better than 1~mK at 2.7 K according to 
calculation.  During calibration the sky acts as a back drop
to the external calibrator, so residual transmission is still nearly
2.73 K radiation. Most of the radiation reflected by the external calibrator
was originally emitted by it and then returned by reflection within the 
concentrator, contributing to the effective emissivity. The next largest source 
of incident radiation towards the calibrator is the emission from the concentrator
but this is also at 2.7 K and thus introduces negligible error. The only 
important sources of radiation on the calibrator not at 2.7 K are the detector 
at 1.65 K and the general instrument at $\sim$1.5 K.  The measured and calculated
reflectance of the calibrator for this radiation is $<3\times 10^{-5}$
at 1 cm$^{-1}$ and diminishes rapidly with frequency. Thus the spectrum
uncertainty of the FIRAS instrument due to the external calibrator is
approximately 10 parts per million.  The full story of the external calibrator
will be covered in a future paper.

The temperatures of the two calibrators and associated concentrators are 
controllable from 2 to 25 K.  Redundant  thermometers measure the 
temperatures of
these four temperature controlled elements and other infrared  emitters 
such as the moving mirrors, the mechanical structure, and the detector 
housings. When observing the sky, the spectrometer is operated with its 
output nearly nulled, by adjusting the internal calibrator temperature.  
This reduces sensitivity to gain errors and instrument drifts.

The path difference $x$ between the two  arms of the interferometer is varied
by scanning a pair of mirrors at a  constant velocity.  This motion produces a
modulated power $P(x)$ at the output  ports. $P(x)$ is approximately the cosine
transform  of the net power
\begin{equation}
P(x) \approx {1\over 2}\int_0^\infty d\nu\,\sum_{\hbox{inputs}} 
E_i(\nu)S_i(\nu) \cos (2\pi\nu\,x),
\end{equation}
where $E_i$ is the effective instrument \'etendue for each emitting element,
$S_i$ is the spectral intensity
of each emitter, and $\nu$ is the frequency in \icm. The principal 
inputs are the sky (or external
calibrator) and the internal calibrator, which are $180\deg$~out of phase
and thus enter into Eqn. 1 with opposite signs. Each of
the outputs  of the interferometer also receives emission from other components
of the instrument.  

The mirrors are scanned at one of two velocities (``slow" and ``fast"), and
over one of two lengths (``short" and ``long"). An insignificant amount of
data was taken in the long slow mode so there are 12 data sets (4 detectors at
each of 3 scan modes). The long data (up to 5.9 cm excursion) were truncated 
to match the short data (with maximum 1.2 cm excursion) for this analysis. 
In retrospect perhaps all of the data should have been taken in the short
slow mode but at the launch of COBE the detailed nature of the CMBR was not
as clear as it is now. The use of two velocities allows clear separation
of some instrument effects (eg. vibration) from optical or sky effects.

Although the high frequency ($\gt 20$ \icm) radiation was effectively excluded
from the low frequency detectors, the filter reflected low frequency radiation 
into the high frequency detectors.  We do not use the 6 high frequency data sets
as the calibration for these is more susceptible to systematic error.
The detectors do not have equal noise. Thus even though we include 5 additional 
data sets we only double the statistical weight of the ``left low short slow" 
data set used in the previous publications. 

The FIRAS points $\approx 94\deg$ from the Sun and away from the Earth
as the {\em COBE} orbits the Earth
with a period of 103~minutes.
The orbit plane is maintained approximately perpendicular to the direction
to the Sun. This scan pattern concentrates the observing
time near the ecliptic poles.
We divide the sky into 6144 equal area pixels according to a
quadrilateralized spherical cube representation (O'Neill and Laubscher 1976).
Each pixel has $\sim0.2$ of the beam area.

\section{Data Reduction}

The data are sorted, ``deglitched", averaged, calibrated, averaged again and
``destriped" by a series of programs.  At each step variance is calculated
and statistics of the results checked.  The final result is a map of 6144
pixels (76 are unobserved) with a spectrum at each pixel.

\subsection{Sorting}
A total of 1411829 good (free from transmission errors) 512 point sky 
interferograms were received. Approximately 20\% of the sky data is rejected 
for various conditions (eg. sky horn temperature, high noise, bolometer 
temperature too high etc). After sorting into the 12 data sets, 11 mission 
periods, 6114 pixels, and several instrument states the remaining 1138149 
interferograms are sorted into 272525 groups with an average of $\sim 4$ 
interferograms per group or coadd. There are an average of 187 interferograms
per pixel corresponding to an integration time of $\sim 30 $ minutes.

In the first part of the mission, the calibrator was commanded into the horn
approximately 2 days each month.  However for the last 6 weeks of the mission
the calibrator was commanded into the horn approximately 3 days each week.
Thus approximately 10\% of the data set is calibration data. 

There are 149260 calibration interferograms. Approximately 10\% of the 
calibration interferograms are rejected.  The fraction is lower for the 
calibration data because it was primarily collected near the end of the 
mission when we had more experience operating FIRAS. The remaining 134739
interferograms are sorted into 3028 coadds with an average of 44 interferograms
per coadd. The calibration coadds are larger because they are not divided into 
pixels.  The maximum size coadd is arbitrarily set at 100.  

\subsection{Deglitching}

A robust estimate of each coadd average is formed at each point in the
interferogram, to aid in recognizing glitches. This average is 
subtracted from each interferogram to yield a ``noise" interferogram.
The individual interferograms are ``deglitched" using a modified 
``CLEAN" algorithm (Isaacman \etal\ 1992). The glitches are recognized by
their shape which is due to the bolometer and amplifier response to
a delta function. Glitches are probably due to cosmic particle hits on the 
detectors so they are not correlated to the position of the mirrors.  Thus 
they add only noise. In the raw data most of the noise is in the glitches so
deglitching is critical to the noise.

We average the center half of the data to form a robust template. 
In the case of $4n$ interferograms, the top $n$ points and the bottom $n$ 
points are discarded and the remaining $2n$ points are
averaged.  In the case of $4n+1$ or $4n-1$ interferograms the top $n$
points and bottom $n$ points are discarded leaving $2n+1$ or $2n-1$ to be 
averaged.  While this is not exactly the center half, it is close and there
is no bias.  In the case of $4n+2$ interferograms our previous analysis ignored 
the bottom $n+1$ points and the top $n$ leaving $2n+1$ points to be averaged. 

Clearly this biases the template, but a simple bias in the interferogram 
domain leads to only a DC term in
the spectral domain, and the DC term (zero frequency) is not used. Since both 
sky and calibration observations are broad band features, the central peak 
of the interferogram has most of the signal and hence some real
signal. This extra signal causes the ``CLEAN" algorithm to infer extra 
``glitches" at the peak only. In the unbiased
case this regresses the data towards the mean, which is a robust average
of the group so this is not a problem.
In the case where the template is biased this induces a small deviation in the 
interferogram, which becomes a small amplitude wide band distortion in the
spectrum.  This is not enough by itself to cause a problem, because the 
calibration data
use the same procedure and thus suffer the same distortion. However, the
calibration data are collected in large groups, while the sky coadds form 
small groups, as the sky data are scattered over many pixels. This bias 
appeared most pronounced in the sky data using only six interferograms.

To fix this bias, we changed the template for the case with $4n+2$ 
interferograms to discard the bottom $n$ points and the top $n$ points. 
The top point and bottom point of the remaining $2n+2$ points are included 
in the average with the other $2n$ points but with only half the weight as a 
better approximation to the average of the center half. 

For groups with fewer than eight interferograms, we included interferograms 
from neighboring pixels (up to 8 total) to form the robust estimate for 
deglitching only. If there were a sufficient number of interferograms (8 or 
more) we did not use neighboring pixels. After the data were deglitched the 
interferograms from the original pixel were averaged like the other groups.  
This borrowing of neighboring interferograms allows us to deglitch data where 
there are only a few interferograms in a group.  This allows us to add an 
additional 15\% of the data and closes many of the gaps that are otherwise 
unobserved. We checked this procedure to verify that it did not introduce 
biases. The noise in a single interferogram is high so
it is possible to use data from a neighboring pixel to approximate
the original data. The neighboring pixels are sampled by the beam which is 
almost 3 pixels across and there is little gradient in most of the sky so the 
actual signal difference is small. We selected the closest interferograms 
(in angle) from neighboring pixels from the same time period and at the same
Galactic latitude for $|b|<20^\circ$. 

The same result is obtained when
we exclude the data using neighboring pixels.  We also compared 
directly the spectra with and without using neighbors in the deglitching.
Finally we compared the spectra with and without using neighbors in the high
frequency Galactic plane data.  This is a more sensitive test, because the
Galaxy exhibits high gradients and sharper spectral features.  In all cases 
the differences were consistent with the expected random noise.

The other features of the deglitching remain as in the previous work (see
Isaacman \etal\ 1992).
As in the previous calibration in some cases a second template was subtracted to
allow for gradients across a pixel and the ``CLEAN" procedure was iterated.

\subsection{Variance Estimation}

Each interferogram is assigned a statistical weight before averaging.
We estimate the variance as a function of glitch rate (or glitch number, 
as the duration of all interferograms is the same for a given scan mode).  
This is modeled as $V=a+bg$, where V is the variance, g is the glitch rate and
a and b are parameters fit for each detector and scan mode. From the deglitching
there are estimates of the variance for each of the 272525 coadds. These 
variance estimates are fit to $V=a+bg$ for each detector and scan mode. The 
weight is then modeled as $1/(a+bg)$.  Although the group estimated variances, 
$v$, are not biased, a weight, which is the inverse of the variance, is biased 
high by the presence of fluctuations. $\langle w_{est}\rangle=\langle 
1/v\rangle~>~1/\langle v\rangle=\langle w_{true}\rangle$. This is particularly
important (25\% effect) for the small coadds of the sky but it is also
significant for the larger coadds of the calibration data. To minimize this 
bias we use an entire data set to fit for $a$ and $b$
rather than the variance of a single group to estimate variance and weight.
Typically 25 glitches halves the weight, 
and there are typically 15 glitches in each interferogram. The weighted average 
interferogram is calculated and converted to a spectrum by apodization, 
zero padding, and Fourier transformation. The total weight of the 
coadd is carried along as a weight of the spectrum. 

The weighted average of the covariance at each frequency was estimated for each 
detector and scan mode using the dispersion of these weighted spectra. This 
variance was used in following calculations to compute the statistical weights. 
When fitting in the frequency domain we need to include the correlation effects.
We use the inverse of the covariance matrix to weight the data.
The covariance between frequencies is well approximated by $C_{\nu\nu'} = 
\sigma_{\nu}~\sigma_{\nu'}~Q(|\nu-\nu'|)$. The expression $|\nu-\nu'|$\ can
take on the values (0,0.4538,...,19.06) for which $Q$ takes on the values
 (1.000, 0.176,$-$0.203, 0.145, 0.077,$-$0.005,$-$0.022, 0.032, 0.053, 0.025,$-$0.003,
 0.007, 0.029, 0.029, 0.003,$-$0.002, 0.016, 0.020, 0.011, 0.002, 0.007, 0.011,
 0.009, 0.003,$-$0.004,$-$0.001, 0.003, 0.003,$-$0.001,$-$0.003, 0.000, 0.003, 0.009, 
 0.015, 0.008, 0.003,$-$0.002, 0.000,$-$0.006,$-$0.006, 0.000, 0.002, 0.008). The
values for $\sigma$ are in Table 3 for the dipole and in Table 4 for the 
CMBR spectrum.

The covariance matrix could also be calculated analytically by propagating 
delta functions through the apodization, Fourier transform, and calibration
process. These generally agree but there are parameters of the electronic
filters, effects of self apodization, and effects of the deglitching
which complicate the calculation so we use the estimated covariance matrix.

\subsection{Calibration}

The \calpaper\ calibration procedure was used.  This includes $\sim 3000$
parameters for each detector and scan mode.  Some of the parameters are nonlinear
and the process involves considerable computation. For a more detailed
description see  \calpaper. Here we will only address the changes in that
procedure. Rather than weighting according
to the variance of each spectrum at each frequency, we used the mission average 
for the noise spectrum shape (dependence on frequency) for each type of 
spectrum (particular setting of temperatures) and the glitch weighting
to scale the noise of each coadd.

The Nyquist frequency (set by the sampling rate) is 145.212 \icm. Our use of a 
320 point spectrum leaves 43 significant elements in the 2 to 21 \icm\ frequency
range.  The excursion of the mirrors is not symmetric with about 1.2 cm in one
direction and 0.5 cm in the other. Thus there is information down to 0.4 cm$^{-1}$
rather than only to 0.56 cm$^{-1}$. By zero padding the short end of the 
interferogram out to a total of 640 points, we can use more of the information 
from the Fourier transform which otherwise winds up in the imaginary part of 
the spectrum. The resulting real elements are not independent but our analyses 
include correlation effects.
 
The truncated long and short interferograms are calibrated together.
There are eight data sets or ``channels" (4 detectors $\times$\ 2 speeds).
The data taken with the external calibrator in the input aperture were used to 
construct a calibration model for each channel.
Each model includes 7 parameters to describe the bolometer, 2 parameters for
the J-FET amplifier, 6 parameters for the spectrometer and 5 parameters for the
unwanted vibrations.  The model is nonlinear in these parameters. The 
temperature 
of each of the four ``controllables" and an overall time shift for each coadd 
are also parameters of the fit, with constraints consistent with the thermometer
readout noise (temperature dependent). The model is almost linear in these 
parameters. A complex number at each frequency parameterizes
the emission from each of seven sources: internal and external calibrators,
sky and reference concentrators, mirrors, structure, and bolometer housing,
with the Kirchhoff constraint that they sum to zero. The model is linear in 
these parameters. The complex number for the external calibrator is 
identified as the transfer function, as the
external calibrator is black and fills the beam.

The $\chi^2$ for the calibration for each channel is given 
in Table \ref{X2}. 
Excess $\chi^2$ indicates an imperfect model of the bolometer and vibration,
and other possible errors. The imperfect bolometer model is also
indicated by preflight data on the bolometers alone. The $\chi^2$/DOF for 
the cold null, sky-like coadds are much closer to unity.  This is expected
because the
hot calibrations stress the model much more than the near null sky conditions.

The fitted calibration models are then applied to both the sky and
calibration spectra, resulting in a sky coadds and calibration coadds in
absolute flux units. After calibration the sky spectra are 
adjusted to the Solar System barycenter to account for the Earth's orbital 
velocity around the Sun, which would otherwise be readily detectable as an 
additional variable dipole of amplitude 0.3 mK. Although this further
correlates the data, the correlation effect is negligible since the Earth's 
velocity around the Sun is $\sim10^{-4}c$. 

\subsection{Destriping}

After the barycenter corrected coadds were produced, a linear fit was made
to the sky and calibration coadds from all four low frequency
channels at each frequency.  The fit includes 14 time dependent
templates, 4 temperature templates, and a parameter for the sky emission
at each pixel and frequency. Thus 18 stripe parameters and 6068 fluxes are 
used to fit $\sim 100000$ coadds separately at each frequency.
By simultaneously fitting all of the data sets with a single ``best" sky 
and a photometric model, we get the maximum amount of cross checking 
on each data set.  By simultaneously fitting the calibration data
we effectively recalibrate the data, get the best sky signal, and get a 
covariance matrix to measure the correlations we have introduced by this 
process. The $\chi^2$ for this fitting is shown in Fig.~\ref{fstripe}.

The templates consist of two global decaying exponentials with time constants
of 61 and 153 days, and step functions for three time periods for each of the
four channels.  The exponentials were included
because an exponential-like decay was seen in the data, both in the sky data
and in the calibration data (Fixsen \etal\ 1994b). 
The cause of this is not known but it is seen in all of the detectors 
and scan modes. The time period templates were
unity during a certain period and zero at other times. Two of the periods were 
when the concentrator temperatures were raised to 6K and 4K respectively. 
The third period covers the rest of the time when the concentrator temperatures 
were 2.73K. The transition between these periods was very short and not used. 
The temperature of the dihedral mirrors was also used as a template because 
parts of the 
instrument had temperature fluctuations of $\sim 2$\ K, correlated with the 
dihedral mirror temperature. We use a separate model for each channel because
the higher temperatures may have different effects on the different detectors.
The calibration corrected for these, but there is 
uncertainty in the correction. We reduce the residual uncertainty by using 
the more abundant sky data to determine the corrections.

\section{The \FIRAS\ Temperature scale}

\specpaper\ gave the absolute temperature of the CMBR as 2.726 $\pm$ 0.010 K, 
with a conservative systematic uncertainty estimate. They noted a discrepancy
between the thermometers and the color temperature scale. We have accounted
for most of the discrepancy by using the low frequency lines to calibrate the
frequency scale rather than the high frequency scale. Since the uncertainty in
going from the FIRAS temperature scale to the CMBR temperature scale in only
a few $\mu$K we discuss them together.  Only the absolute temperature of the
external calibrator is an issue since the other parts of the instrument
only need to be stable. There are three ways of 
determining the CMBR temperature from the FIRAS data set:

\begin{enumerate}
\item
We use the preflight calibration of the external calibrator thermometers, 
which should be good to the nominal 1 mK accuracy of the calibration 
specification. 
The task is quite simple since there exist substantial
calibration data with the external calibrator at 2.73 K and other temperatures
near the sky conditions.  This gives a CMBR temperature of $2.730\pm0.001$~K, 
with the error entirely dominated by the absolute thermometry calibration 
error on the external calibrator.  The main 
difficulty with this is that there is evidence that the internal calibrator 
thermometry has errors of $\sim 3$ mK and that the thermometer at 
the tip of the external calibrator differs from the other two by 3 mK 
(at 2.73 K).
There is some evidence from testing the alternate calibrators on the ground
that both of these problems may be due to insufficient thermal contact for
the thermometers embedded in epoxy at the tip, and that the two thermometers 
on the external calibrator fixed to the copper base ring should not have this 
problem.  These were used in estimating the external calibrator temperature
and the tip temperature is not used.

\item
We use the calibration data to determine the temperature scale.
This effectively uses the ``color" to determine the temperature and so depends 
on the frequency determination. We have improved the frequency determination 
of the \FIRAS\ (see Fig~\ref{ffreq}).  While \specpaper\ use 
the C$^+$ (63 \icm) and N$^+$ (49 \icm) lines, we note the possibility that
the high frequency calibration and the low frequency calibration need not agree.
A discrepancy can arise because the beam in the interferometer has divergence
(5.5\deg HWHM on average) which because of the cosine effect implies a 0.5\% 
frequency shift. However the beam profile of the high frequency light may be 
more strongly attenuated than 
the beam profile of the low frequency data.  
To improve the frequency scale we used 
a combination of CO and C$^+$ lines (7.69, 11.53, 15.38, and 16.42 \icm) to 
make 
our low frequency determination to $\pm0.03$\%.  The difference between the 
high frequency and low frequency scales implies a difference of 1\deg in 
average beam divergence. The 0.03\% frequency uncertainty implies a temperature 
uncertainty of 0.82 mK. There is an additional 0.2 mK error in determining the 
color temperature once the frequency scale is set, but as this adds in 
quadrature it is negligible.  The result of this analysis is that the CMBR 
temperature is $2.7255 \pm 0.0009$ K.

\item
We can also use the CMBR itself.  If we assume the dipole is a result of 
a Doppler shift the shape of the differential spectrum should be 
$dB_\nu/dT$, where $B_\nu(T)$ is the Planck function. The best fit temperature 
to the dipole spectrum (see below) is a CMBR temperature of $2.717 \pm 0.007$ K.
This result depends on a differential spectrum so offsets of the instrument 
calibration do not matter. Since the 
measurements are taken 50 minutes apart (on opposite sides of the orbit), long
term drifts (of the sort we see and correct for) also do not matter.
This measurement is not completely independent because it still depends on
the frequency calibration. The uncertainty is dominated by the uncertainty in 
fitting the Galaxy radiation which modulates the dipole signal which is only 
0.1\% of the CMBR signal in the Rayleigh-Jeans region.

\end{enumerate}

These three methods give answers that are within three
sigma.  Apparently there are additional systematic errors in one or
more of these methods.  We conclude the absolute temperature of 
the CMBR is $2.728\pm 0.004$\ K (95\% CL), entirely dominated by the systematic 
errors. While this is not a true statistical uncertainty it is a 
useful summary of the uncertainty in the result.

\section{Analysis of Sky Data}

The calibrated destriped sky spectra were then fit to four spatial templates. 
We use the FIRAS data to determine the spectra of the four components.
We model the data $S(\nu;\ell,b)$, where $\ell, b$ are Galactic coordinates and 
$\nu$ is frequency, as follows:
\begin{equation}
S(\nu;\ell,b) = I_\circ(\nu) + D(\ell,b)d(\nu) 
+ G_1(\ell,b)g_1(\nu) + G_2(\ell,b)g_2(\nu)\quad.
\end{equation}
The monopole is represented by the spectrum $I_\circ(\nu)$; the
dipole variation is represented by the spatial distribution $D(\ell,b)$ and the
spectrum $d(\nu)$; and the Galactic emission is represented by one or two 
spatial distributions $G_k(\ell,b)$ and corresponding spectra $g_k(\nu)$.
The fit is made independently at each frequency, only the spatial variation is
assumed.  Each of the resulting spectra will be discussed and further modeled
below.  In all cases the uncertainties due to this template fit are propagated
to the following analysis.

To make this separation the functions $D(\ell,b)$ for the dipole and
$G_k(\ell,b)$ for the Galactic emission must be specified.  The dipole is
$D(\ell,b) = \cos(\theta)$, where $\theta$ is the angle between the observation
and the maximum of the dipole, 
$(\ell,b)=(264.26^\circ,+48.22^\circ)$ 
(Bennett \etal\ 1996). 

For $G_k(\ell,b)$ we tried several combinations of predictors. 
Five templates for $G(\ell,b)$ are:

\begin{enumerate}

\item
A plane-parallel, $\csc|b|$ distribution.  This is clearly an 
oversimplification of the Galaxy, but our results are independent of the 
Galactic template used. This provides strong evidence that our results are not 
contaminated by any inadequacy of the model of the Galactic emission.

\item
The spatial distribution of the total power received in the high frequency 
\FIRAS\ channel above 25~\icm. This is used under the assumption that the high 
frequency ($25<\nu<80\,$cm$^{-1}$) radiation is well correlated to the low 
frequency ($2 <\nu<21\,$cm$^{-1}$) Galactic radiation. The high frequency 
channel is centered on 41~cm$^{-1}$ with a 44~cm$^{-1}$  effective
bandwidth. Since the high and low frequency data are taken simultaneously 
through the same concentrator there is little mismatch in sky coverage and 
no need for beam convolution. 

\item
\COBE\ DIRBE (Diffuse Infrared Background Experiment) 240 $\mu$m map,
convolved to \FIRAS\ resolution. This has the advantage of being totally 
independent and low noise but could suffer from beam convolution errors.
The DIRBE resolution is $\sim 0.7^\circ$ and the FIRAS resolution is 
$\sim 7^\circ$.

\item
\COBE\ DIRBE 140 $\mu$m map,
convolved to \FIRAS\ resolution. This in combination with the 240 $\mu$m
allows adjustments for varying dust temperatures.

\item
\COBE\ DIRBE 100 $\mu$m map,
convolved to \FIRAS\ resolution. This has lower noise than the 140 $\mu$m
data but it is further afield in frequency and may be contaminated by other
effects (zodiacal emission).

\end{enumerate}

For Galaxy templates 2-5 we use the normalization $\langle G(\ell,b)\rangle_{
|b|>60^\circ}=1.074$, the natural normalization of the csc$|b|$ model.
The only effect of the normalization, of course, is to rescale the Galaxy spectrum.
With this normalization the resulting Galaxy spectrum is the sky brightness
near the Galactic poles.

The principal fit for this paper uses the DIRBE 240 and DIRBE 140 combination 
for $G_k(\ell,b)$ and a 5\deg~Galactic cut. There is a high $\chi^2$ ``tail"
(see Fig 1) so after the fit was made, pixels 
with a $\chi^2/$DOF$>2$ were removed.  The fit was repeated with the remaining 
pixels and again pixels with a $\chi^2/$DOF$>2$ were removed.This 
process was repeated until the pixels removed formed a stable set.  This 
removes 58 pixels.

To investigate the sensitivity to these parameters many other fits were made.
We also tried Galactic cuts of 0, 10, 15, 20, 30, and 40 degrees. 
We tried fits to subsets of the data. We also tried each of the five Galactic 
templates individually and the DIRBE 240 \& DIRBE 100, DIRBE 240 \& high 
frequency FIRAS and high frequency FIRAS \& DIRBE 140 combinations.
While one spatial Galactic template gives a reasonable result a proper 
combination of two templates greatly reduces the dependence on the Galactic cut.
We also varied the $\chi^2/$DOF cutoff from 1.3 to 3.
For other Galactic models or other Galactic cutoffs of 
course different pixels were removed.  In particular for single component
Galactic models many pixels at low Galactic latitudes were removed.

These other fits form a background ``cloud" of solutions which test the result
for systematic effects. We use the RMS deviation of a solution amongst these
to estimate the size of the systematic uncertainties.

The data are weighted according to the 
channel variance including glitch effects (see section 3.3). The spectra which 
are determined by the fit are the linear coefficients of the three or four 
spatial
components discussed above, made independently at each frequency.  The 
result can be viewed as three or four spectra, one for each spatial component.
It is 
important to note that this fit makes no a priori assumptions about the forms 
of these spectra; only their spatial distribution is posited.

\subsection{Galactic Spectrum}

To get a single Galactic spectrum we add the Galactic spectra if there are 
more than one. The result is the average spectrum near the Galactic poles.
The Galactic spectrum can be approximately fit by functions of the form 
$\nu^n B_{\nu}(T_{\rm dust})$. Wright \etal\ (1991) examined the \FIRAS\ 
Galaxy spectrum and found that it was fit by $n = 1.65$ and $T_{\rm dust} = 
23.3$~K. The Galactic spectrum derived here, which is determined by high 
latitude Galactic data, produces a lower temperature, $T_{\rm dust}=13 \pm 1$
~K, with $n$ fixed at 2 and an emissivity of $\sim 10^{-4}$ at 60 cm$^{-1}$.
With only the low frequency data used here we cannot fit for both $n$ and
the temperature. 
About half of temperature difference is due to the change in index from 1.65 
to 2 and half from the change to the higher latitudes. Even emphasizing the 
high Galactic latitude regions the CO and C$^+$ lines are still evident.

When using the DIRBE 240 and DIRBE 140 templates to fit the Galactic radiation 
we note that the DIRBE 140 template produces a spectrum that is negative for 
all frequencies, however the strong correlation between DIRBE 240 and DIRBE 140 
ensures that the sum is positive everywhere.  

The Galactic fit to $\csc|b|$ has more radiation at the Galactic poles than
the fit to the other Galactic models, indicating that the Sun is in a
local ``hole" in the Galaxy.  This local ``hole" is approximately 2/3 as dense
as a typical Galactic location.

The dust can be modeled with other forms as well. Most of the information on
the Galactic regions is at higher frequencies.
A full discussion of the Galaxy requires the high frequency data
(see Bennett 1994 and Reach 1995);
here the Galactic radiation is treated as a contaminant of the CMBR.

\subsection{Dipole Spectrum}

We fit the dipole spectrum to the derivative of a Planck function
$T_{\rm amp}dB_\nu(T_{\rm cmbr})/dT$ where we fix $T_{\rm cmbr} = $\tcbr\ 
and allow $T_{\rm amp}$ to vary.
The best fit value of the dipole amplitude is $T_{\rm amp} =3.369$ mK $\pm
0.004$ mK. However the value is dependent on the Galactic cutoff and the Galaxy 
model.  In our ``cloud" of solutions we find an RMS variation of 0.006 mK.  
Adding this in quadrature
gives an uncertainty of 0.007 mK. We also make a 0.003 mK correction for the 
finite beamsize, which is approximately $7^\circ$ FWHM. The result is 
summarized in Table~\ref{tderr}. This amplitude of the dipole is in agreement 
with $3.353\pm.024$ mK measured by the {\em COBE} DMR (Bennett \etal\ 1996), 
derived from observations at 1.05, 1.77, and 3.00 cm$^{-1}$. The DMR beam is
also approximately 7$^\circ$ but the details of the beam patterns are different.
We do not include possible effects of the octupole and higher harmonics 
although because of our nonuniform sky coverage these might perturb our result. 
Bennett \etal\ 1996 includes these effects.

The the dipole spectrum and the fitted Planck derivative
are shown in Figure~\ref{fDipRes} and in Table~\ref{tDipRes}. The weighted
deviations, $\sqrt{\sum_\nu{(\frac{\Delta_\nu}{\sigma_\nu} )^2}
/\sum_\nu \frac{1}{\sigma_\nu^2}}$, from the blackbody derivative dipole 
spectrum are \dipdev\ rms, smaller than the deviations of the monopole
from its fitted blackbody, primarily because it is less susceptible to 
systematic errors. 

By choosing the monopole temperature as the point to evaluate $dB_\nu/dT$, we
have forced the dipole temperature to be that of the monopole.
If we fit both $T_{\rm amp}$ and $T_{\rm cmbr}$
we obtain $T_{\rm cmbr} = $\tdip$\pm$0.003~K
($1\,\sigma$ statistical uncertainty).
However the result is sensitive to the Galactic cut.  We add a systematic 
uncertainty of 0.006 to account for this effect.
This results in a final estimation of  $T_{\rm cmbr} = $\tdip$\pm$0.007~K
($1\,\sigma$ combined uncertainty).
The agreement of the dipole temperature with the monopole temperature is
consistent with the conventional Doppler interpretation of the dipole.
It is also a confirmation that the overall temperature scale and 
calibration for the FIRAS are consistent. 

We can also find the dipole direction from the FIRAS data with a different fit.
Here we reverse the assumptions of the principal fit. We assume three spectra,
a CMBR temperature and two Galactic dust optical depths, one for each of the 
spectra derived from the DIRBE templates in the principal fit, $g_k(\nu)$, and 
fit for their intensities at each pixel. The CMBR temperature assumes a Planck 
spectrum. These spectra were 
chosen to match the shape of the spectrum in the data.
This yields maps of the CMBR temperature and dust intensities. A monopole 
plus three dipole components are then fit to the resulting temperature map.

The vector sum of the dipole coefficients points in the direction 
\dipdir, consistent with the direction 
from the DMR results.  Data for $|b|<10^\circ$ were excluded from the dipole 
fit because of the potential inaccuracy of the model of the Galaxy. The 
direction is particularly sensitive to the galaxy because it is almost
orthogonal to the direction to the Galactic center. Galactic variations in 
spectral shape as a function of longitude couple into the angle directly while 
for the fixed angle case they come in as second order terms for the same reason.

\section{The Cosmic Spectrum}

The monopole spectrum (see Fig 4) is well fit by a Planck blackbody  spectrum, and
deviations are small, consistent with the earlier \FIRAS\ results within 
their larger uncertainties (Shafer \etal\ 1992, Cheng \etal\ 1992, and
Mather \etal\ 1994).  To determine or constrain any deviations  from a 
blackbody, consider a generic cosmological model $S_c(\nu;p)$, where $p$  is 
some cosmic parameter quantifying the deviation from a blackbody,  such as the 
Kompaneets $y$ parameter for Comptonized  spectra or the dimensionless chemical 
potential $\mu$ for a  Bose-Einstein photon  distribution.  Because the 
deviation is small, a linear fit
\begin{equation}
I_0(\nu) = B_\nu(T_0) + \Delta T {{\partial B_\nu}\over{\partial 
T}}
             + G_0 g(\nu) + p {{\partial S_c}\over{\partial p}} 
\label{withgal}
\end{equation}
can  be performed to the unknown parameters $p$, $G_0$, and $\Delta T$. The
first two terms are the Planck blackbody spectrum, with the  temperature  $T_0
+ \Delta T$.  The third term allows for Galactic contamination remaining in 
the monopole spectrum.  The final term is the modeled deviation. We fit either
the Kompaneets parameter or the chemical potential, but the two are too 
similar to fit them simultaneously. The uncertainties are propagated from the
template fits, and the correlation between the $g(\nu)$ and 
$ {{\partial S_c}\over{\partial p}}$ increases the uncertainty of $G_0$
and $p$.
 
\subsection{Galactic contamination}

We use either the $\sum_k g_k(\nu)$ derived from the all sky data set 
(see section 4) or $\nu^2B_\nu(T)$ and fit a temperature and
emissivity. The $\nu^2B_\nu(T)$ model produces a lower $\chi^2$ with a 
temperature of 9 K and
we use this model for the analysis in Table 4.  One possible interpretation
of this fit is a 9 K Galactic halo or a cosmic background (uniform component) 
with a spectrum similar to the Galactic spectrum. One must be cautious however
as subtle variations in dust temperature or emissivity can produce similar 
effects.

The determination of $g(\nu)$ is dominated by low latitude Galactic emission,
and 
there is some variation from this form at higher latitudes. We vary the
Galactic latitude cutoff used in deriving $I_0(\nu)$ in eqn. (3) to test the effect
of variations in the Galactic spectrum from $g(\nu)$. Variations greater than
the statistical uncertainty in  any derived parameters, such as the
cosmological term $p$, would most likely be due to an inadequacy in our
Galactic model. Indeed if the $\chi^2$ cutoff is raised and the csc$|b|$ model 
is used such effects can be seen.

When $p$ is fixed at 0, \ie there is only a Galactic deviation fit, the 
weighted rms deviation in the frequency range 2~to 21~\icm\ is \rmspct\ 
(parts per million) of the peak brightness.  The weighted 
rms is defined as $\sqrt{\sum_\nu (\frac{\Delta_\nu}{\sigma_\nu})^2 / \sum_\nu 
\frac{1}{\sigma_\nu^2}}$, where $\Delta_\nu$ is the difference of the model 
and the data at spectrum point $\nu$, and $\sigma_\nu$ is the uncertainty of 
the spectrum point $\nu$. The formal $\chi^2$ per degree of freedom = 46/40 
= 1.15. The residuals, with the errors, are shown in Fig.~\ref{ftres} and 
listed in Table~\ref{ttres}.  The best fit temperature is \tcbr, which is
determined by our choice of frequency and thermometry scales (see section 4).
When comparing this to values from other experiments, the systematic 
thermometry uncertainty of \tcbrerr\ ($1\sigma$) should be included.
The detector noise part of the temperature uncertainty is only \trand, and is 
entirely negligible.  

\subsection{Bose-Einstein Distortion}

There are two likely forms for distortions of the primeval spectrum produced
during the plasma epoch before recombination, and we compute a limit
independently for each. Energy release or conversion in
the redshift range $10^5 < z < 3\times 10^6$ produces a Bose-Einstein
distribution, where the Planck law is modified by  a dimensionless chemical 
potential $\mu$ (Zel\'dovich \& Sunyaev 1970):
\begin{equation}
S_\mu(\nu;T,\mu) = {{2hc^2\nu^3} \over {e^{x+\mu} - 1}},
\label{eSmu}
\end{equation}
where $x = hc\nu / kT$
and $\nu$ is measured in \icm. The linearized deviation of $S_\mu$ from a 
blackbody is the derivative of Eq. 4 with respect to $\mu$:
\begin{equation}
{\partial S_\mu \over \partial \mu} = {{-T_0}\over{x}}{{\partial B_\nu 
}\over{\partial T}}\quad.
\end{equation}

The current \FIRAS\ result is  $\mu = $\muval, or a 95\%\ confidence level upper
limit  of $|\mu| < $\mumin.  This result and those following are 
summarized in Table 2, including the effects of systematic uncertainties 
and the variation of the result according to the Galactic model used in the fit.

\subsection{Compton Distortion}

Energy release at later times, $z < 10^5$, produces a  Comptonized spectrum, a
mixture of blackbodies at a range of temperatures.  In the case  of
non-relativistic electron temperatures this spectrum is described by  the
Kompaneets (1957) equation, parameterized by the value of $y$ (Zel\'dovich \&
Sunyaev 1969):
\begin{equation}
y = \int {{k(T_e-T_\gamma)}\over{m_ec^2}} d\tau_e\quad,
\end{equation}
where $T_e$, $T_\gamma$ and $\tau_e$ are the electron temperature, the  CMBR
photon temperature, and the optical depth to electron Compton scattering. The
distortion will be of the form (Zel\'dovich \& Sunyaev 1969)
\begin{equation}
{{\partial S_y}\over{\partial y}} = T_0 \Bigl( x {\rm coth}(x/2)
 - 4 \Bigr) {{\partial B_\nu }\over{\partial T}}\quad.
\end{equation}
The results are $y =$\yval. There is some dependence on Galactic cut and on the 
Galactic model.  This indicates either residual Galactic contamination
or a nearly uniform source with a spectrum simular to the Galactic spectrum.
We include a systematic uncertainty of $4\times 10^{-6}$ to account for these 
effects for a final 95\%\ confidence level upper limit of \ymin.

The limit for $|\mu|$ is about 1/3 as large as that given in \specpaper\ and
the limit on $|y|$ is about 1/2 as large.
The $|y|$ is more strongly coupled to the residual Galactic contamination.
These new values place stringent constraints on theories of the early universe
and the development of cosmic structure.   

\section{Interpretation}

The cosmological interpretation of the limits on $y$ and $\mu$ are  developed
by Wright \etal\ (1994).  We summarize their conclusions  here briefly. Since
the \FIRAS\ spectrum with the Galactic dust emission removed is  close to a
blackbody, strict limits can be set on energy release  after the
Big Bang.  For energy release during the Bose-Einstein period,  the limit is
$\Delta U / U = 0.71 \mu $ where $U$ is the energy  of the cosmic background
and $\Delta U$ is the energy converted from  other forms.  For energy release 
after that but before decoupling, the limit is $\Delta U / U = 4y$. The result
can be summarized as a limit on $\Delta U/U$ as a function of redshift
shown in Fig 6. 

\section{Summary and Conclusions} 

The FIRAS spectrum of the cosmic microwave background radiation agrees with a
blackbody spectrum to high accuracy.  The CMBR monopole and dipole spectra are 
the result of fitting a model including a dust map derived from the DIRBE
data. The CMBR dipole has a spectrum consistent with its thermal origin and a 
Doppler shift.  The dipole itself has a differential thermal spectrum the 
temperature of which, \tdippm, agrees with the monopole temperature.
The Doppler shift implies that the Sun's peculiar velocity relative to the 
comoving frame is $371\pm1$ km/s (95\% CL) toward \dipdir, 
in agreement with the microwave results from the DMR.

The CMBR temperature is \tcbrpm, where the error is dominated by our 
estimate of the thermometry errors.  The weighted rms deviation from the fit 
is \rmspct\ of the peak brightness.  The limit on $|y|$ is 
\ymin\ and the limit on $|\mu|$ is \mumin\ (95\% CL).

\section{Acknowledgments} 

We thank the computer scientists and analysts who operated the \COBE,
processed the data, and solved problems in software that could no longer be
solved with hardware.  In particular, we thank S.~Alexander, K.~Jensen,
S.~Read, F.~Shuman, and A.~Trenholme. This work was supported by the 
Astrophysics Division of NASA Headquarters.

\clearpage
\singlespace

\begin{table}
\begin{center}
\begin{tabular}{lrrrr}
& \multicolumn{2}{c}{Low Frequency}& \multicolumn{2}{c}{High Frequency}\\
& Left& Right& Left& Right\\
\tableline
Short Slow & 68143 & 53795 & 451509 & 558325 \\
DOF        & 34856 & 35276 & 205650 & 200216 \\
\tableline
Fast	   & 20785 & 17198 & 228365 & 229018 \\
DOF        & 13688 & 14360 &  96552 &  95298 \\
\end{tabular}\\
\end{center}
\caption{Calibration $\chi^2$ and number of degrees of freedom for each 
channel.}
\label{X2}
\end{table}

\begin{table}
\begin{center}
\begin{tabular}{lrrrr}
& Fit & Statistical& Systematic& Final\\
& Value& uncertainty& uncertainty& uncertainty\\
\tableline
Galaxy Temp & 13.3  & 0.6  & 0.8  & 1.0 K \\
Dipole Amp  & 3.372 & 0.004 & 0.006 & 0.007 mK \\
Dipole Temp & 2717 & 3 & 6 & 7 mK \\
Gal Latitude & 48.26 & 0.11 & 0.10 & 0.15 deg\\
Gal Longitude & 264.14 & 0.14 & 0.06  & 0.15 deg\\
CMBR Temp & 2.728 & 0.00001 & 0.002 & 0.002 K \\
$y$ para & -1 & 6 & 4 & 7 $\times 10^{-6}$ \\
$\mu$ para & -1 & 4 & 1 & 4 $\times 10^{-5}$ \\
\end{tabular}\\
\end{center}
\caption{Results of fits.}
\label{tderr}
\end{table}

\begin{table}
\begin{center}
\begin{tabular}{rrrr}
Frequency &  Measurement & Residual & Uncertainty \\ [-2ex] 
\icm\ & &kJy/sr & 1$\sigma$ \\ [-.5ex]
\tableline \\ [-5ex]
  2.27 &   458& 33&  7\\ [-2ex]
  2.72 &   605& 23&  9\\ [-2ex]
  3.18 &   770& 23& 12\\ [-2ex]
  3.63 &   956& 45& 11\\ [-2ex]
  4.08 &  1106& 38& 11\\ [-2ex]
  4.54 &  1218&  5& 10\\ [-2ex]
  4.99 &  1346&  5&  9\\ [-2ex]
  5.45 &  1473& 26&  9\\ [-2ex]
  5.90 &  1534&  5&  8\\ [-2ex]
  6.35 &  1588&  1&  7\\ [-2ex]
  6.81 &  1623&  2&  6\\ [-2ex]
  7.26 &  1645& 12&  6\\ [-2ex]
  7.71 &  1619& -3&  5\\ [-2ex]
  8.17 &  1576&-17&  5\\ [-2ex]
  8.62 &  1541& -7&  5\\ [-2ex]
  9.08 &  1490&  1&  6\\ [-2ex]
  9.53 &  1423&  2&  7\\ [-2ex]
  9.98 &  1331&-12&  8\\ [-2ex]
 10.44 &  1234&-26&  9\\ [-2ex]
 10.89 &  1171& -3& 11\\ [-2ex]
 11.34 &  1077&-10& 11\\ [-2ex]
 11.80 &   996& -4& 11\\ [-2ex]
 12.25 &   910& -5& 11\\ [-2ex]
 12.71 &   814&-18& 11\\ [-2ex]
 13.16 &   780& 27& 11\\ [-2ex]
 13.61 &   687&  8& 10\\ [-2ex]
 14.07 &   595&-14& 10\\ [-2ex]
 14.52 &   551&  7&  9\\ [-2ex]
 14.97 &   492&  7&  9\\ [-2ex]
 15.43 &   435&  5& 10\\ [-2ex]
 15.88 &   388&  8& 10\\ [-2ex]
 16.34 &   326& -9& 11\\ [-2ex]
 16.79 &   318& 24& 13\\ [-2ex]
 17.24 &   275& 18& 14\\ [-2ex]
 17.70 &   230&  5& 15\\ [-2ex]
 18.15 &   224& 28& 16\\ [-2ex]
 18.61 &   172&  2& 16\\ [-2ex]
 19.06 &   162& 14& 17\\ [-2ex]
 19.51 &   157& 30& 20\\ [-2ex]
 19.97 &   113&  3& 27\\ [-2ex]
 20.42 &   102&  7& 43\\ [-2ex]
 20.87 &   161& 79& 76\\ [-2ex]
 21.33 &    45&-25&139\\ [-4ex]
\end{tabular}
\end{center}
\caption{Dipole spectrum and residuals.  Intensities in kJy/sr. }
\label{tDipRes}
\end{table}

\begin{table*}
\begin{center}
\begin{tabular}{rrrr}
Frequency &  Residual & Uncertainty & Galaxy\\ [-2ex] 
\icm\ & kJy/sr& 1$\sigma$ &at Pole\\ [-.5ex]
\tableline \\ [-5ex]
  2.27&      5& 14&  4\\ [-2ex]
  2.72&      9& 19&  3\\ [-2ex]
  3.18&     15& 25& -1\\ [-2ex]
  3.63&      4& 23& -1\\ [-2ex]
  4.08&     19& 22&  3\\ [-2ex]
  4.54&    -30& 21&  6\\ [-2ex]
  4.99&    -30& 18&  8\\ [-2ex]
  5.45&    -10& 18&  8\\ [-2ex]
  5.90&     32& 16& 10\\ [-2ex]
  6.35&      4& 14& 10\\ [-2ex]
  6.81&     -2& 13& 12\\ [-2ex]
  7.26&     13& 12& 20\\ [-2ex]
  7.71&    -22& 11& 25\\ [-2ex]
  8.17&      8& 10& 30\\ [-2ex]
  8.62&      8& 11& 36\\ [-2ex]
  9.08&    -21& 12& 41\\ [-2ex]
  9.53&      9& 14& 46\\ [-2ex]
  9.98&     12& 16& 57\\ [-2ex]
 10.44&     11& 18& 65\\ [-2ex]
 10.89&    -29& 22& 73\\ [-2ex]
 11.34&    -46& 22& 93\\ [-2ex]
 11.80&     58& 23& 98\\ [-2ex]
 12.25&      6& 23&105\\ [-2ex]
 12.71&     -6& 23&121\\ [-2ex]
 13.16&      6& 22&135\\ [-2ex]
 13.61&    -17& 21&147\\ [-2ex]
 14.07&      6& 20&160\\ [-2ex]
 14.52&     26& 19&178\\ [-2ex]
 14.97&    -12& 19&199\\ [-2ex]
 15.43&    -19& 19&221\\ [-2ex]
 15.88&      8& 21&227\\ [-2ex]
 16.34&      7& 23&250\\ [-2ex]
 16.79&     14& 26&275\\ [-2ex]
 17.24&    -33& 28&295\\ [-2ex]
 17.70&      6& 30&312\\ [-2ex]
 18.15&     26& 32&336\\ [-2ex]
 18.61&    -26& 33&363\\ [-2ex]
 19.06&     -6& 35&405\\ [-2ex]
 19.51&      8& 41&421\\ [-2ex]
 19.97&     26& 55&435\\ [-2ex]
 20.42&     57& 88&477\\ [-2ex]
 20.87&   -116&155&519\\ [-2ex]
 21.33&   -432&282&573\\ [-4ex]
\end{tabular}
\end{center}
\caption{Residual of monopole
spectrum.  Intensities in kJy/sr.  }
\label{ttres}
\end{table*}

\clearpage
\typeout{FIGURE CAPTIONS}
\begin{figure}
\caption{$\chi^2$/DOF in the destriping process.
a) $\chi^2$/DOF as a function of frequency for low frequency channels.
b) $\chi^2$ distribution for pixels for low frequency data.}
\label{fstripe}
\end{figure}

\begin{figure}
\vspace {.8in}
\caption{
Frequency of $C^+$ line as a function of Galactic longitude.  Galactic rotation
is clearly evident. Diamond shows low frequency measurement, which was used to 
set the frequency scale.}
\label{ffreq}
\end{figure}

\begin{figure}
\vspace {.8in}
\caption{Dipole spectrum and fit to $\frac{dB}{dT}$. Vertical lines indicate
plus and minus one $\sigma$ uncertainties. Peak of uniform CMBR is 
approximately 400 MJy/sr.} 
\label{fDipRes}
\end{figure}

\begin{figure}
\vspace {.8in}
\caption{Uniform spectrum and fit to $Planck(T)$. Uncertainties are 
a small fraction of the line thickness.} 
\label{fPlanck}
\end{figure}

\begin{figure}
\vspace {.8in}
\caption{FIRAS measured CMBR residuals,(------)
$I_0 - B_\nu(T_0) - \Delta T \frac{dB}{dT} - G_0 g(\nu)$.
Spectrum model components:
the maximum allowed distortions (95\% CL)
$y = $ \ymin\ (-- -- --) and $|\mu| = $ \mumin\ ($\cdots$);
the Galaxy spectrum $g(\nu)$ scaled to one 
fourth the flux at the Galactic pole ($\cdot\;$--), and the effects of a 
100~$\mu$K  
temperature shift in $T_0$, $0.0001\* \frac{dB}{dT}$, 
($\cdots\;$--).}
\label{ftres}
\end{figure}

\begin{figure}
\vspace {.8in}
\caption{Derived limit on injected energy as a function of redshift.}
\label{fdeltaU}
\end{figure}


\clearpage
\begin{references}

\reference Bennett, C.~L. \etal, 1992a, ``Recent Results from COBE," in {\it 
The Evolution of Galaxies and Their Environment}, H.A. Thronson \& J.M. Shull, 
eds., Proc. Third Teton Summer School

\reference Bennett, C.~L. \etal, 1992b, \apj, 396, L7

\reference Bennett, C.~L. \etal, 1994, \apj, 436, 423

\reference Bennett, C.~L. \etal, 1996, In preparation

\reference Boggess, N.~W. \etal, 1992, \apj, 397, 420

\reference
Cheng, E.~S. 1992, ``Far-Infrared Cosmology Measurements,'' in 
Proceedings of the International Symposium on Observational Cosmology, Milan, 
Italy, September 1992

\reference Fixsen, D.~J. \etal, 1994a, \apj, 420, 445

\reference Fixsen, D.~J. \etal, 1994b, \apj, 420, 457

\reference Isaacman, R., Read S., \& Barnes, W. 1992, in {\it Astronomical Data
Analysis Software and Systems I} ed. D. M. Worall, C. Biemesderfer, \& J. 
Barnes, (Astronomical Society of the Pacific Conf. Ser., Vol. 25) p. 403

\reference Kompaneets, A. S.  1957, JETP, 4, 730

\reference Mather, J. C. \etal, 1990, \apjl, 354, L37

\reference Mather, J.C., Fixsen, D.J. and Shafer, R.A. ``Design for the COBE
    Far  Infrared Absolute Spectrophotometer (FIRAS)," COBE Preprint
    93-10, Proc. SPIE, vol 2019, pp. 168-179, conf. on Infrared
    Spaceborne Remote Sensing, in San Diego, CA, 11-16 July 1993,
    (SPIE: Bellingham, WA)

\reference Mather, J. C. \etal, 1994, \apj, 420, 440

\reference O'Neill, E. M. \& Laubscher, R. E. 1976,
Extended Studies of a Quadrilateralized Spherical Cube Earth Data Base,
(NEPRF Technical Report 3-76)
(NTIS Report AD-A026294)
(Computer Sciences Corp. CSC/TR-76/6008)

\reference Peebles, P.~J.~E. 1971, {\it Physical Cosmology}, (Princeton:
Princeton University Press)

\reference Reach, W. T. \etal, 1995, \apj, 451, 188

\reference Shafer, R. A., \etal, 1992, \baas, 23, 1398

\reference Sunyaev, R. A. \& Zel'dovich, Ya. B.  1980, \araa, 18, 537

\reference Wright, E. L. \etal, 1991, ApJ, 381, 200

\reference Wright, E.~L. \etal, 1994, \apj, 420, 450

\reference Zel\'dovich, Ya. B. \& Sunyaev, R. A. 1969, \apss, 4, 301

\reference Zel\'dovich, Ya. B. \& Sunyaev, R. A. 1970, \apss, 7, 20

\end{references}
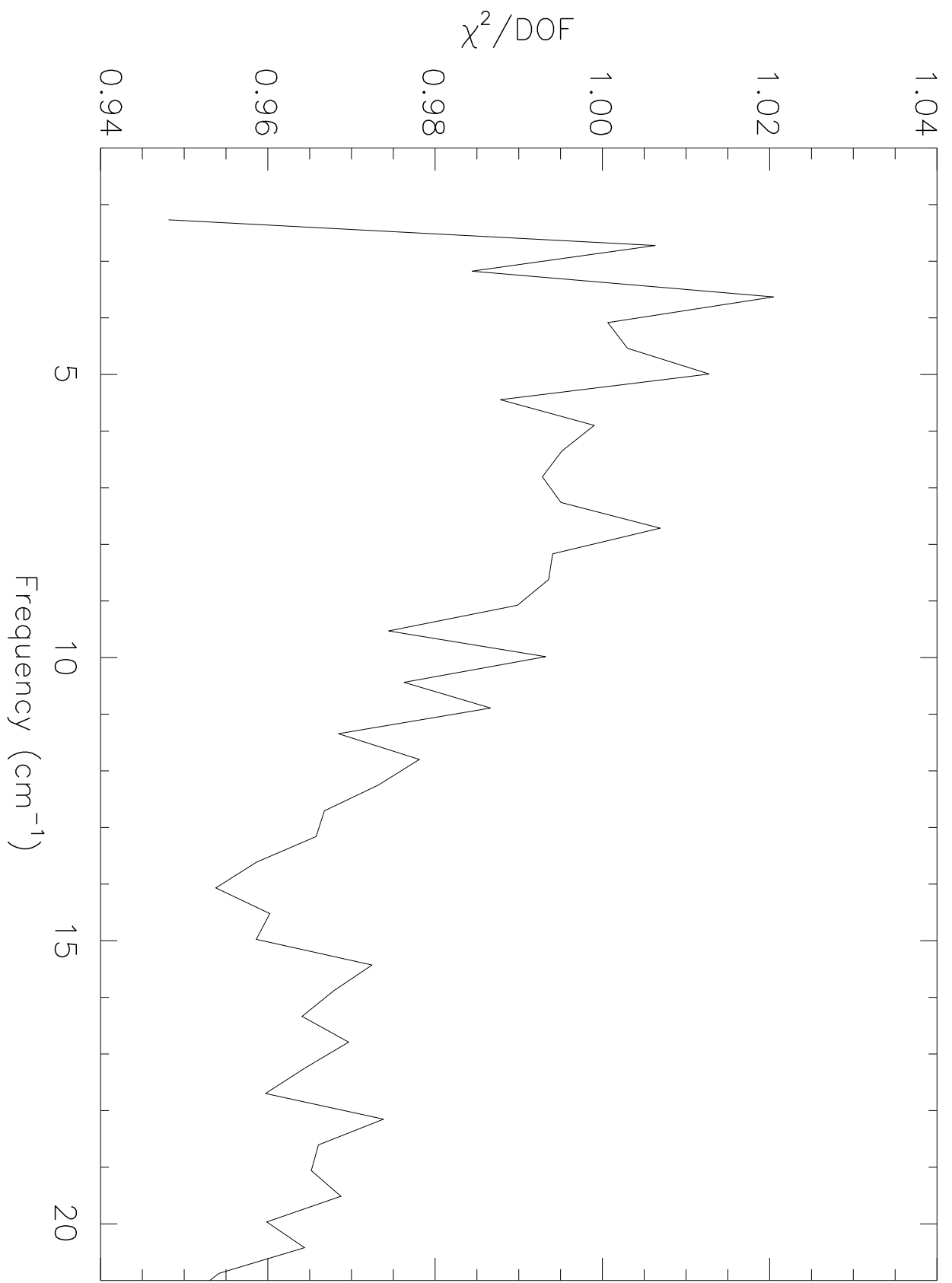
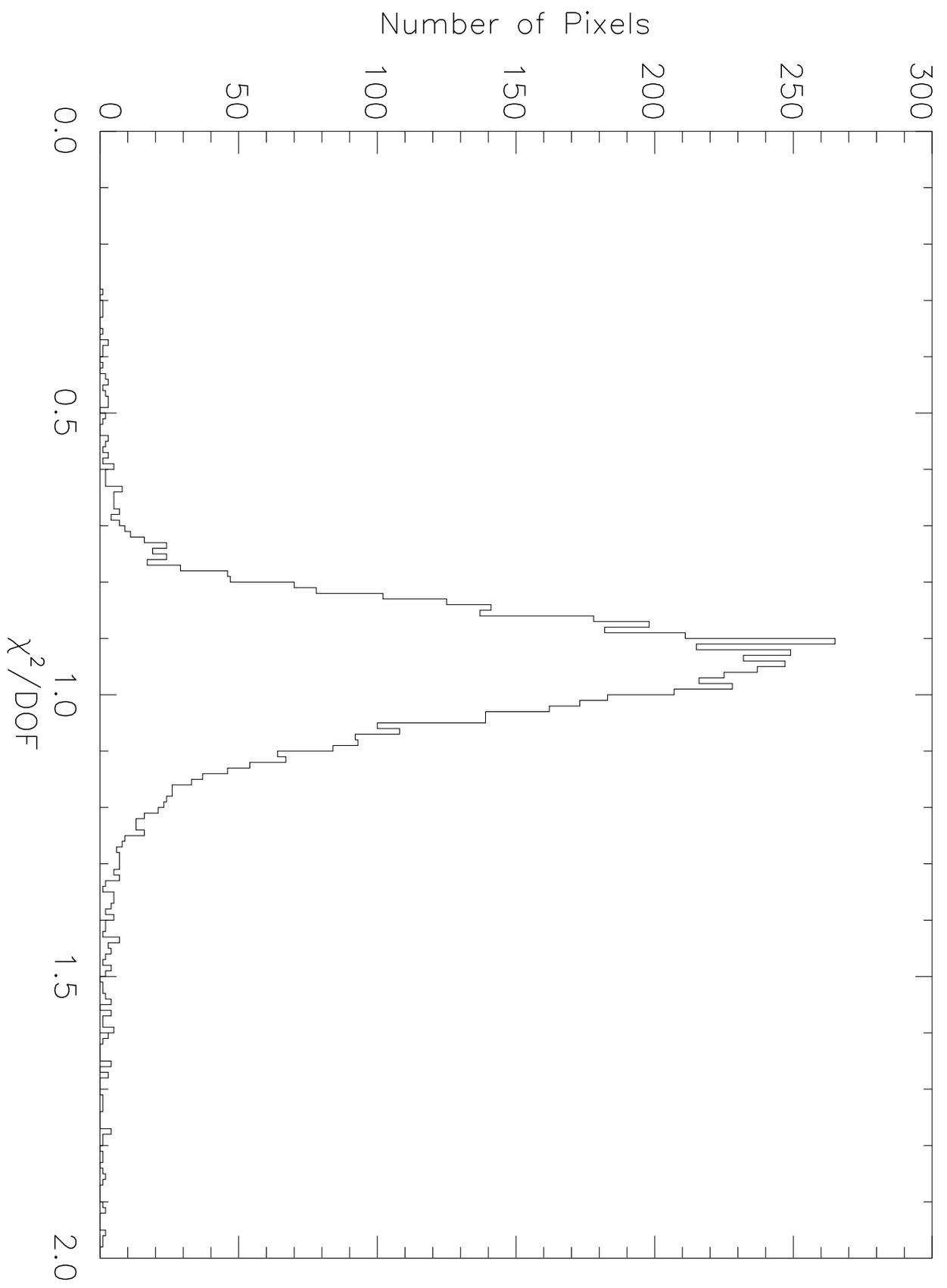
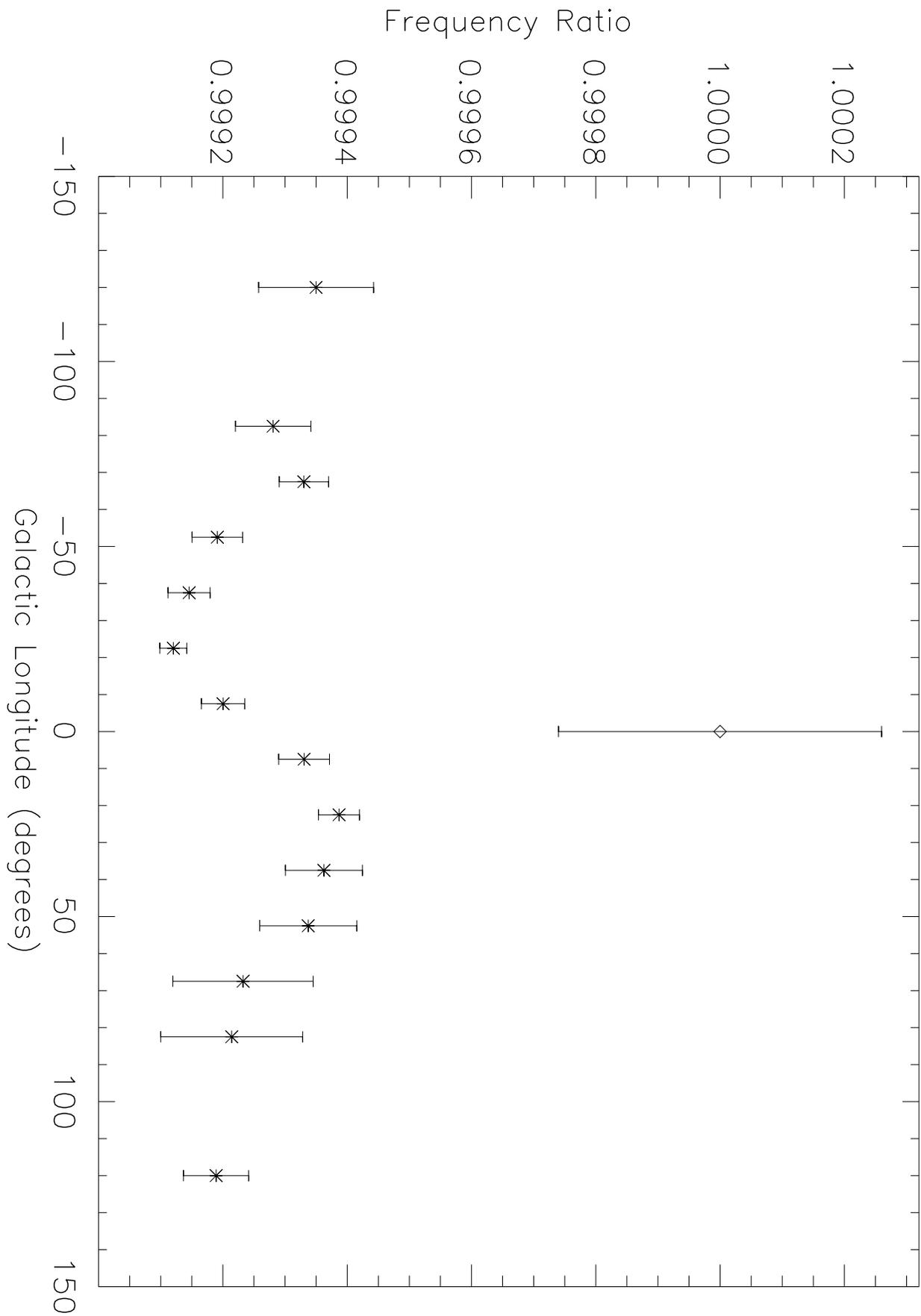
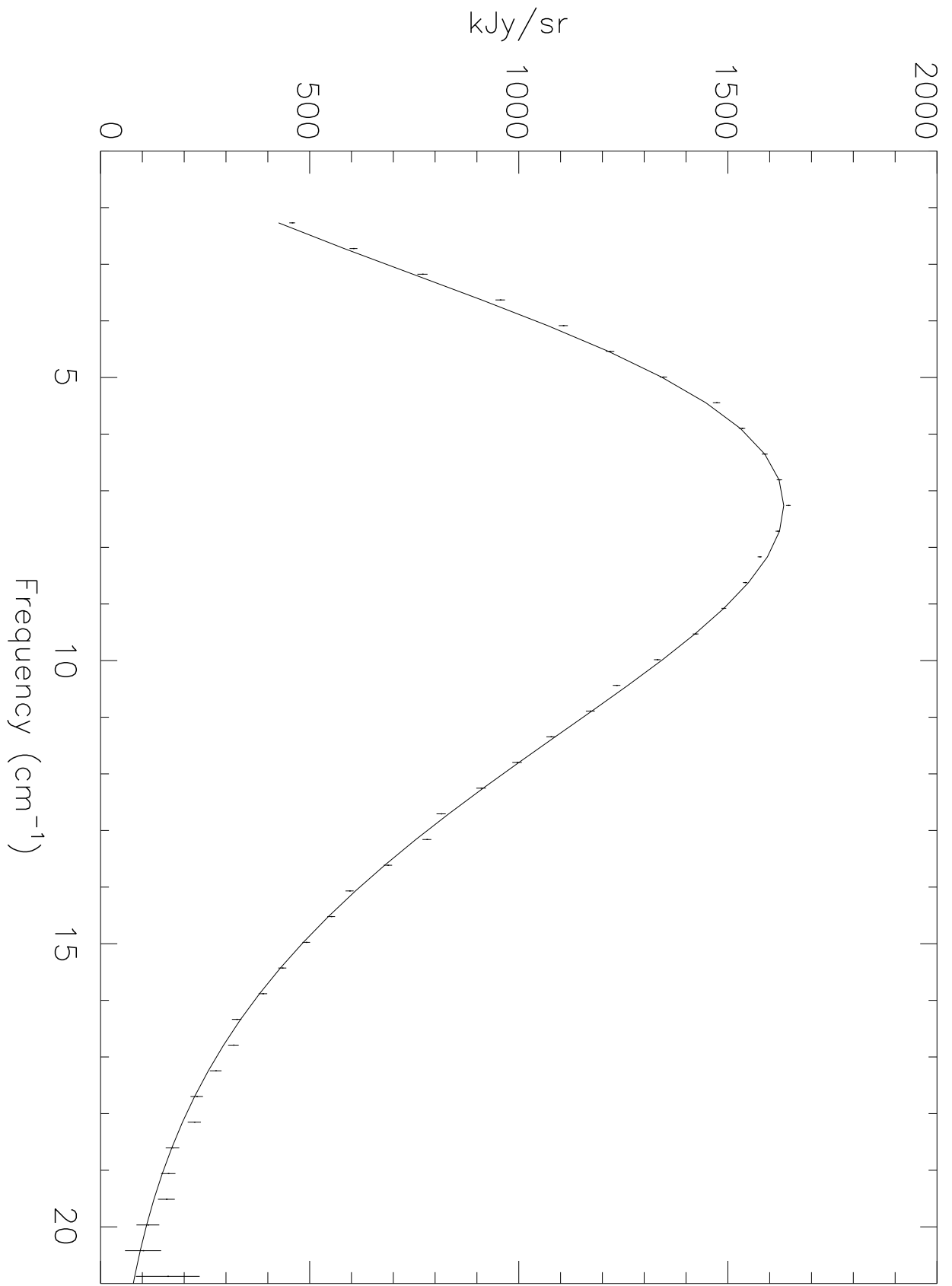
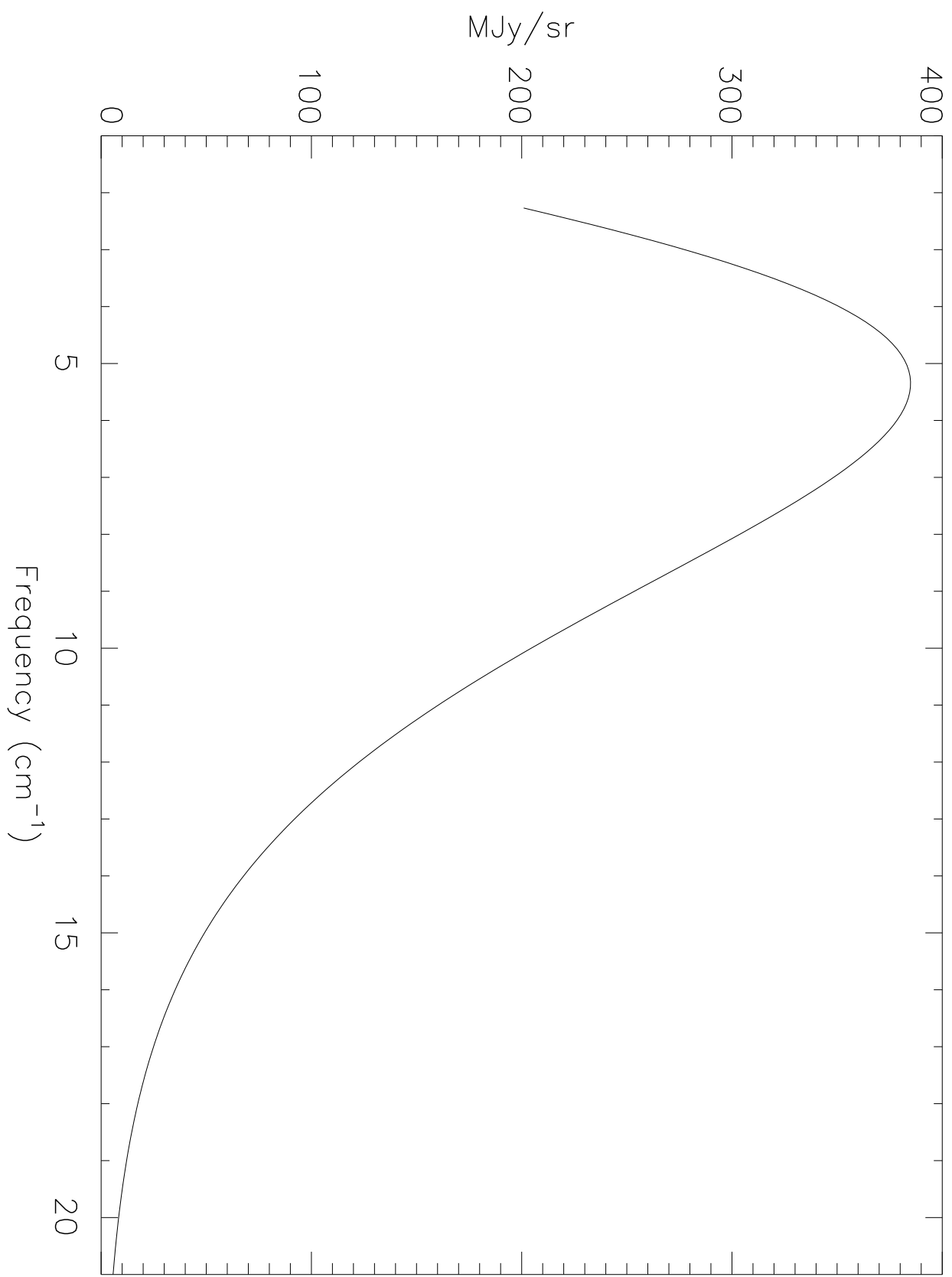
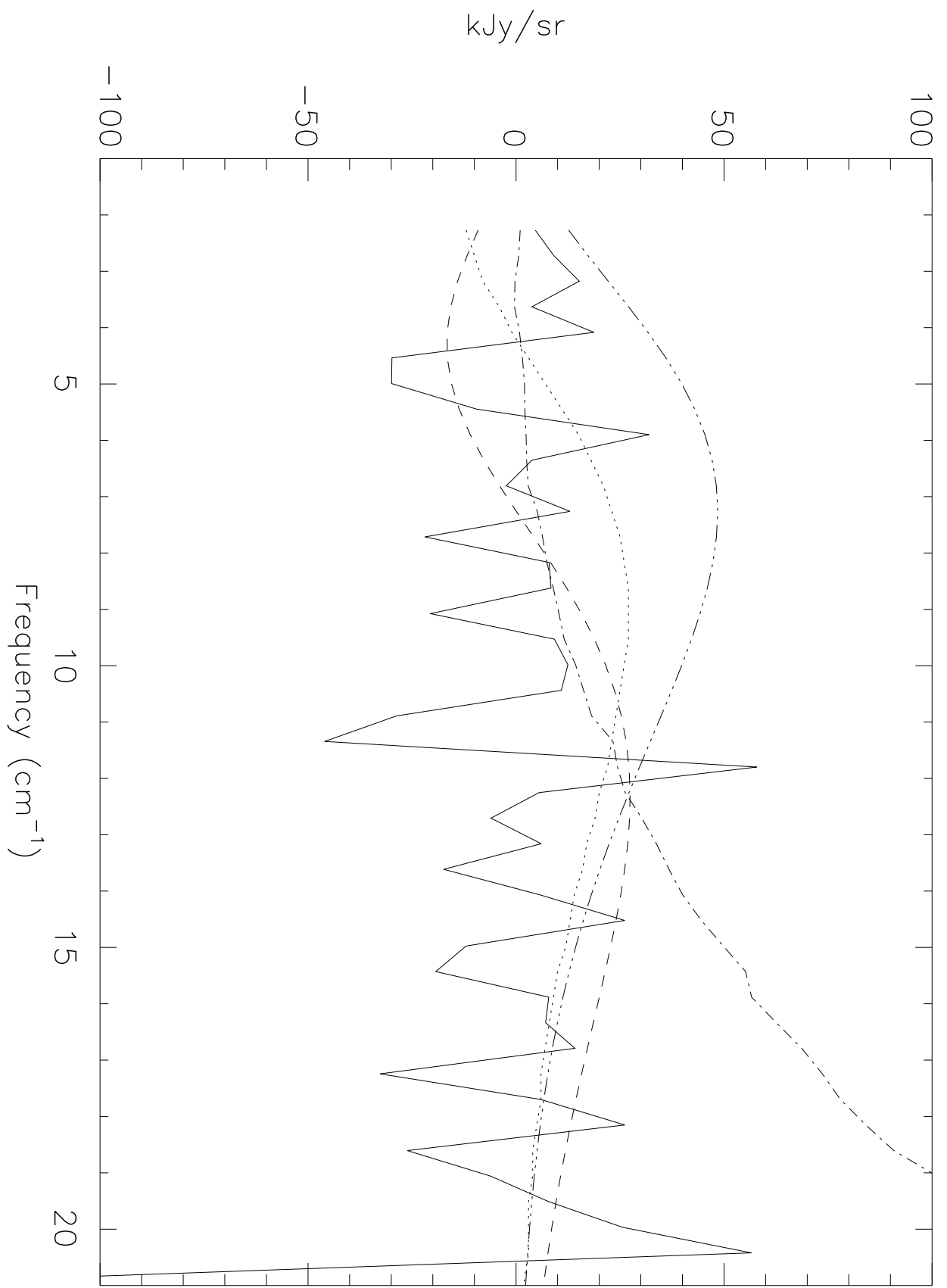
\end{document}